\documentstyle[12pt]{article}
\begin{document}
\begin{center}
{\Large \bf  Lagrangians and Hamiltonians 
 for High School Students}
\end{center}
\begin{center}
{\large John W. Norbury}\\
{\em Physics Department and Center for Science Education,
University of Wisconsin-Milwaukee,
P.O. Box 413, Milwaukee, Wisconsin 53201, USA
e-mail:  norbury@uwm.edu}
\end{center}

\centerline{{\em Abstract}}

A discussion of Lagrangian and Hamiltonian dynamics is
presented at a level which should be suitable for advanced high school students. This is intended for
those who wish to explore a version of mechanics beyond the usual Newtonian treatment in high
schools, but yet who  do not have advanced mathematical skills.

\section{Introduction}
Newtonian dynamics is usually taught in high school physics courses and in
college level freshman physics class \cite{1}.  Lagrangian and Hamiltonian
dynamics \cite{2,3} is usually reserved for an upper division undergraduate
physics course on classical dynamics.  This is all as it should be,
particularly since one needs the technique of calculus of variations for
the Lagrangian formulation.

	However it is always nice to be able to whet the appetite of the advanced
high school student for a taste of things to come.  For those students who
have successfully mastered the contents of the typical high school physics
course, one can give an extra lesson on Lagrangian and Hamiltonian dynamics
without having to use calculus of variations.  The idea is simply to
present some new formulations of dynamics that an advanced high school
student will find enjoyable and intellectually interesting.  (The students
can be told that a rigorous formulation will be presented in college courses.)

For simplicity, consider only the one-dimensional problem.  Write Newton's
equation
\begin{equation}
F=ma \label{1}
\end{equation}
and define the potential energy $U(x)$, which is a function only of
position, as
\begin{equation}
F\equiv -\frac{dU}{dx}\label{2}
\end{equation}
where $-\frac{dU}{dx}$ is the spatial derivative of the potential energy.
Thus re-write Newton's equation as
\begin{equation}
-\frac{dU}{dx} = m\ddot x \label{3}
\end{equation}
where $\dot x \equiv \frac{dx}{dt}=v$ for the speed and $\ddot x \equiv
\frac{d^2x}{dt^2}=a$ for the acceleration.

\section{Lagrangian Dynamics}

To introduce Lagrangian dynamics define a {\em Lagrangian} as a function of
the two variables of position $x$ and speed $\dot x$
\begin{equation}
L(x, \dot x)\equiv T(\dot x)-U(x)=\frac{1}{2} m\dot x^2 - U(x) \label{4}
\end{equation}
where the kinetic energy $T(\dot x)\equiv\frac{1}{2} m{\dot x}^2$ is a
function only of the speed variable and the potential energy again is only
a function of position $U(x)$.

	Now introduce the idea of a partial derivative.  This is very easy.  For a
function of a single variable $f(y)$ the notation $\frac{df}{dy}$ is used
for the derivative.  For a function of {\em two} variables $g(y,z)$ there
are two possible derivatives for each variable $y$ or $z$.  In this case
one simply introduces a different notation for derivative, namely
$\frac{\partial g}{\partial y}$ for the $y$ derivative (where $y$ is
changing but $z$ is constant) and $\frac{\partial g}{\partial z}$ for the
$z$ derivative (where $z$ is changing but $y$ is constant).  Even though
high school students won't see partial derivatives until they are in
college, nevertheless the idea is very simple and can easily be explained
to the advanced student who is taking a course in calculus.

	From (\ref{4}) one can easily see that
\begin{equation}
\frac{\partial L}{\partial x} = -\frac{dU}{dx} \label{5}
\end{equation}
and
\begin{equation}
\frac{\partial L}{\partial\dot x} = m{\dot x}\equiv p \label{6}
\end{equation}
which is called the momentum $p$.  Obviously then
\begin{equation}
\frac{d}{dt} \left( \frac{\partial L}{\partial\dot x}\right) = m{\ddot x}
\label{7}
\end{equation}
Combining (\ref{5}), (\ref{7}) and (\ref{3}), Newton's equation (\ref{3})
becomes
\begin{equation}
\frac{\partial L}{\partial x} = \frac{d}{dt} \left( \frac{\partial
L}{\partial\dot x}\right) \label{8}
\end{equation}
which is the Euler-Lagrange equation in one dimension.  It can be explained
to the students that it is this equation in Lagrangian dynamics which
replaces $F=ma$ in Newtonian dynamics.

\subsection{Lagrangian example}

Students will obviously want to see some examples of how the Lagrangian
formulation works.  A simple example is the one-dimensional harmonic
oscillator with
\begin{equation}
F\equiv -\frac{dU}{dx} =-kx . \label{9}
\end{equation}
Newton's equation is
\begin{equation}
-kx = m\ddot x . \label{10}
\end{equation}
The potential $U(x)$ is obtained by integrating (\ref{9}) to give
\begin{equation}
U(x)=\frac{1}{2} kx^2 . \label{11}
\end{equation}
Thus the Lagrangian is
\begin{equation}
L(x, \dot x) = \frac{1}{2} m\dot x^2 - \frac{1}{2} kx^2 \label{12}
\end{equation}
giving
\begin{equation}
\frac{\partial L}{\partial x} =-kx \label{13}
\end{equation}
and substituting into (\ref{8}) and (\ref{7}) gives exactly back the same
equation of motion (\ref{10}) as in the Newtonian case.

	Many teachers will have had the students work out the equation of motion
from Newtonian dynamics for other types of forces, such as a particle in a
uniform gravitational field.  Students can be encouraged to prove that the
same equations of motion result from the Lagrangian formulation.  Students
can also be encouraged to think about three-dimensional problems and to
derive, on their own, the three Euler-Lagrange equations (corresponding to
the three component equations $F_x=m\ddot x$, $F_y = m\ddot y$, $F_z=m\ddot
z$) which result from the three dimensional Lagrangian
\begin{equation}
L(x,y,z,\dot x, \dot y, \dot z) = \frac{1}{2} m(\dot x^2+\dot y^2+\dot z^2)
- U(x,y,z) .\label{14}
\end{equation}

\section{Hamiltonian Dynamics}

Now consider the Hamiltonian formulation of dynamics.  Define a {\em
Hamiltonian} as a function of the two variables, momentum $p$ and position
$x$,
\begin{equation}
H(p,x)\equiv p\dot x - L(x, \dot x)\label{15}
\end{equation}
which can be seen to be just the total energy $T+U$ as $H=p\dot x - L=m\dot
x^2-\frac{1}{2} m\dot x^2 + U = \frac{1}{2} m\dot x^2 + U = T+U$.
Hamilton's equations follow immediately.  $L$ is not a function of $p$ and
therefore
\begin{equation}
\frac{\partial H}{\partial p} = \dot x . \label{16}
\end{equation}
But $L$ is a function of $x$ and thus
\begin{equation}
\frac{\partial H}{\partial x} =- \frac{\partial L}{\partial x} .\label{17}
\end{equation}
However (\ref{6}) and (\ref{8}) give $\frac{\partial L}{\partial x} = \dot
p$ so that
\begin{equation}
- \frac{\partial H}{\partial x} = \dot p .\label{18}
\end{equation}
Equation (\ref{16}) and (\ref{18}) are Hamilton's equations which replace
$F=ma$ in Newtonian dynamics.

\subsection{Hamiltonian example}

For the harmonic oscillator example, the Hamiltonian is
\begin{equation}
H(p,x) = p\dot x - \frac{1}{2} m\dot x^2 + \frac{1}{2} kx^2 =
\frac{p^2}{2m} + \frac{1}{2} kx^2 \label{19}
\end{equation}
where we have had to replace $\dot x$ by $\frac{p}{m}$ because $H(p,x)$ is
supposed to be a function of $p$ and $x$ only.  Thus Hamilton's equations
(\ref{16}) and (\ref{18}) give
\begin{equation}
\frac{p}{m} = \dot x \label{20}
\end{equation}
and
\begin{equation}
-kx = \dot p . \label{21}
\end{equation}
These are shown to give the equation of motion (\ref{10}) by
differentiating (\ref{20}) as
\begin{equation}
\frac{\dot p}{m} = \ddot x \label{22}
\end{equation}
and substituting (\ref{21}) for $\dot p$ gives back equation (\ref{10}).

	Once again students can be encouraged to use other examples that they have
already studied in Newtonian dynamics and to show that Hamilton's equations
result in the same equation of motion.  Again students can work out the
three-dimensional generalization of Hamilton's equations using
\begin{equation}
H(p_x, p_y, p_z, x, y, z) = p_x\dot x + p_y\dot y + p_z\dot z - L(x, y, z, \dot x, \dot y,
\dot z). \label{23}
\end{equation}
Finally teachers can emphasize to students that Newtonian mechanics is
based on forces, whereas Lagrangian and Hamiltonian dynamics is based on
energy.

	In summary, a discussion of Lagrangian and Hamiltonian dynamics has been
presented which should be suitable for advanced high school students, who
are interested in exploring some topics not normally presented in the high
school physics curriculum.  It is also hoped that this article can be given
to students to read on their own.

\end{document}